\documentclass[letter,traditabstract]{aa}  

\usepackage{graphicx}
\usepackage{txfonts}

\newcommand{\Msun}{\mbox{$M_{\odot}$}}
\newcommand{\lsun}{\mbox{$L_{\odot}$}}
\newcommand{\Lsun}{\mbox{$L_{\odot}$}}

\newcommand{\Zsun}{\mbox{$Z_{\odot}$}}

\begin{document}

   \title{Exploring the Red Supergiant wind kink}

   \subtitle{A Universal mass-loss concept for massive stars}

   \author{Jorick S. Vink
              \and
               Gautham N. Sabhahit
          \inst{1}
          }

   \institute{Armagh Observatory and Planetarium,
              College Hill, BT61 9DG, Armagh, Northern Ireland\\
              \email{jorick.vink@armagh.ac.uk}
  }

   \date{Received August 25, 2023; accepted September 14, 2023}

 
  \abstract
   {Red supergiants (RSG) are key objects in studying the evolution of massive stars and their endpoints, but uncertainties related to their underlying mass-loss mechanism have stood in the way of an appropriate framework for massive star evolution thus far. In this work, we analyse a recently uncovered empirical mass-loss "kink" feature and we highlight its similarity to hot star radiation-driven wind models and observations at the optically thin-to-thick transition point. We motivate a new RSG mass-loss prescription that depends on the Eddington factor, $\Gamma,$ (including both a steep luminosity, $L$, dependence and an inverse steep mass, $M_{\rm cur}$, dependence). We subsequently implement this new RSG mass-loss prescription in the stellar evolution code MESA. We find that our physically motivated mass-loss behaviour naturally reproduces the Humphreys-Davidson limit without the need for any ad hoc tweaks. It also resolves the RSG supernova "problem." We argue that a universal behaviour that is seen for radiation-driven winds across the HR diagram, independent of the exact source of opacity, is a key feature of the evolution of the most massive stars.}

   \keywords{stars: mass loss -- stars: evolution -- stars: supergiants -- stars: massive
               }

   \maketitle
%

\section{Introduction}

Massive stars are those with initial zero-age main sequence (ZAMS) masses upwards of 8\,\Msun. Up to a maximum mass somewhere in the 15-30\,\Msun\ range, they are thought to produce core-collapse supernovae (SNe; e.g., Smartt 2015), but the exact boundaries of neutron star versus black hole formation are still highly uncertain (e.g., Langer 2012; Smith 2014; Meynet et al. 2015; Adams et al. 2017). Related questions involve the red supergiant (RSG) "problem" (Smartt 2015; Kochanek 2020) as well as the physics and metallicity ($Z$) dependence of the Humphreys-Davidson (HD) limit (Humphreys \& Davidson 1979; Davies et al. 2018; Higgins \& Vink (2020); Gilkis et al. 2021; Sabhahit et al. 2021). 

While there is general consensus that hot-star winds are radiation-driven on gas and are expected to depend on $Z$ (Puls et al. 2008; Vink 2022), the situation with respect to the wind-driving of cool RSG winds and their potential $Z$-dependence is still far from clear (e.g., Decin 2021). This holds massive consequences for our lack of understanding of massive star evolution and their final endpoints. 

Beasor et al. (2020) published a RSG mass-loss recipe with rates that were mostly on the low end, but Yang et al. (2023) published rates that were higher. Independently of the absolute rates published (and where discrepancies still need to be addressed), we 
wish to draw attention to a notable feature in the Yang et al. study involving a "knee" or "kink" feature in the mass-loss rate when it is plotted versus luminosity, $L$. A similar "kink" feature was uncovered by Vink et al. (2011) for hot-star winds, where this was attributed to multiple scattering in winds driven by radiation pressure (Vink et al. 2011; Vink \& Gr\"afener 2012). 

Here, we suggest that the RSG kink may also be the result of the onset of multiple scattering at the optically thin-to-thick transition, where $\eta = \tau = 1$ from Vink \& Gr\"afener (2012) in radiation-driven winds of RSGs.
The consequence of a dependence on $L/M$ or the Eddington factor, $\Gamma$ may involve a runaway. 

Gr\"afener \& Vink (2016) suggested such a $\Gamma$-dependent {\it superwind} for the first Flash spectrum of the type IIb SN 
2013cu. And while alternative mechanisms, such as pulsations (Heger et al. 1997; Yoon \& Cantiello 2010) or turbulence (Kee et al. 2021) may also showcase an increasing mass-loss rate when $L$ and/or $\Gamma$ increases, for those mass-loss mechanisms there is no a priori reason to expect a kink. The increased multi-scattering due to a $\Gamma$-dependent wind would be expected to give just that. It is probably also relevant to mention that the predicted hot-star kink of Vink et al. (2011) was empirically verified with observations (Bestenlehner et al. 2014).

\section{A new RSG wind prescription}

Figure~1 shows the observational small magellanic cloud (SMC) data from RSG by Yang et al. (2023). 
These authors fit their data with a third-order polynomial that only depends on $L$, but the key feature of a radiation-driven wind is its dependence on $\Gamma$, namely, proportionally to $L$ and inversely with $M$. 
Indeed, Beasor et al. (2020) did provide a formula that depends steeply on $L$ and inversely on $M$, accounting for the initial masses derived from overall cluster properties. 
We would expect that any type of luminous star that loses significant amounts of mass, to increasingly showcase stronger mass-loss rates, providing a positive feedback loop.

\begin{figure}
    \includegraphics[width=0.49\textwidth]{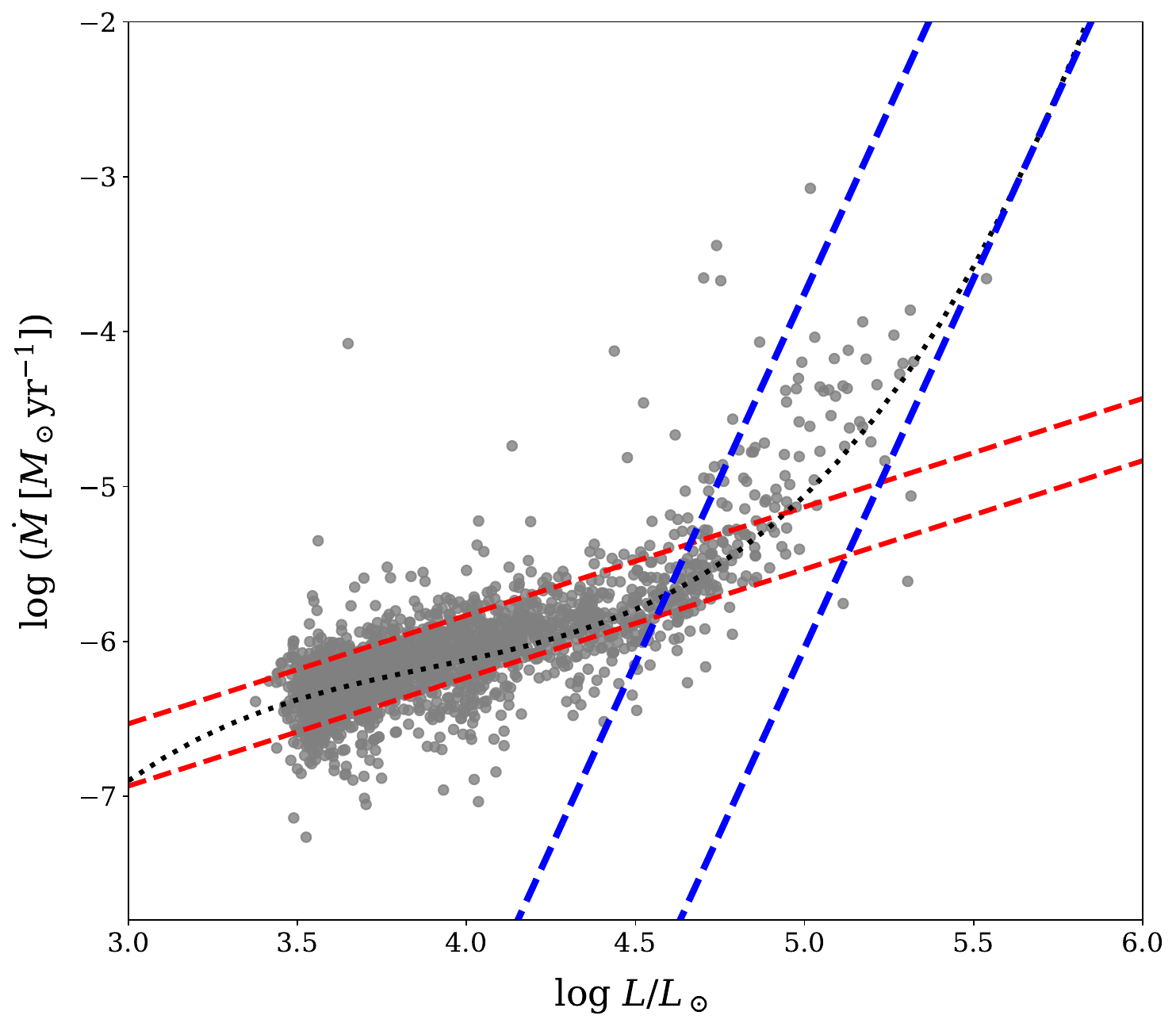}
    \caption{Observational mass-loss rates of Yang et al. (2023) show a kink at log$L/\lsun = 4.6$. We adopt a $\Gamma$ dependent mass-loss relation from our understanding of radiation-driven winds by utilising a shallow and a steep mass loss relation respectively below and above the kink from Vink et al. (2011). 
The two blue and two red lines are for different values of current masses 8\,\Msun\ (upper line) and 30\,\Msun\ (lower line). 
Also shown (black dotted) is a third degree polynomial fit in luminosity derived by Yang et al. (2023).}
    \label{fig:mdot}
\end{figure}

   \begin{figure}
   \centering
   \includegraphics[width=0.48\textwidth]{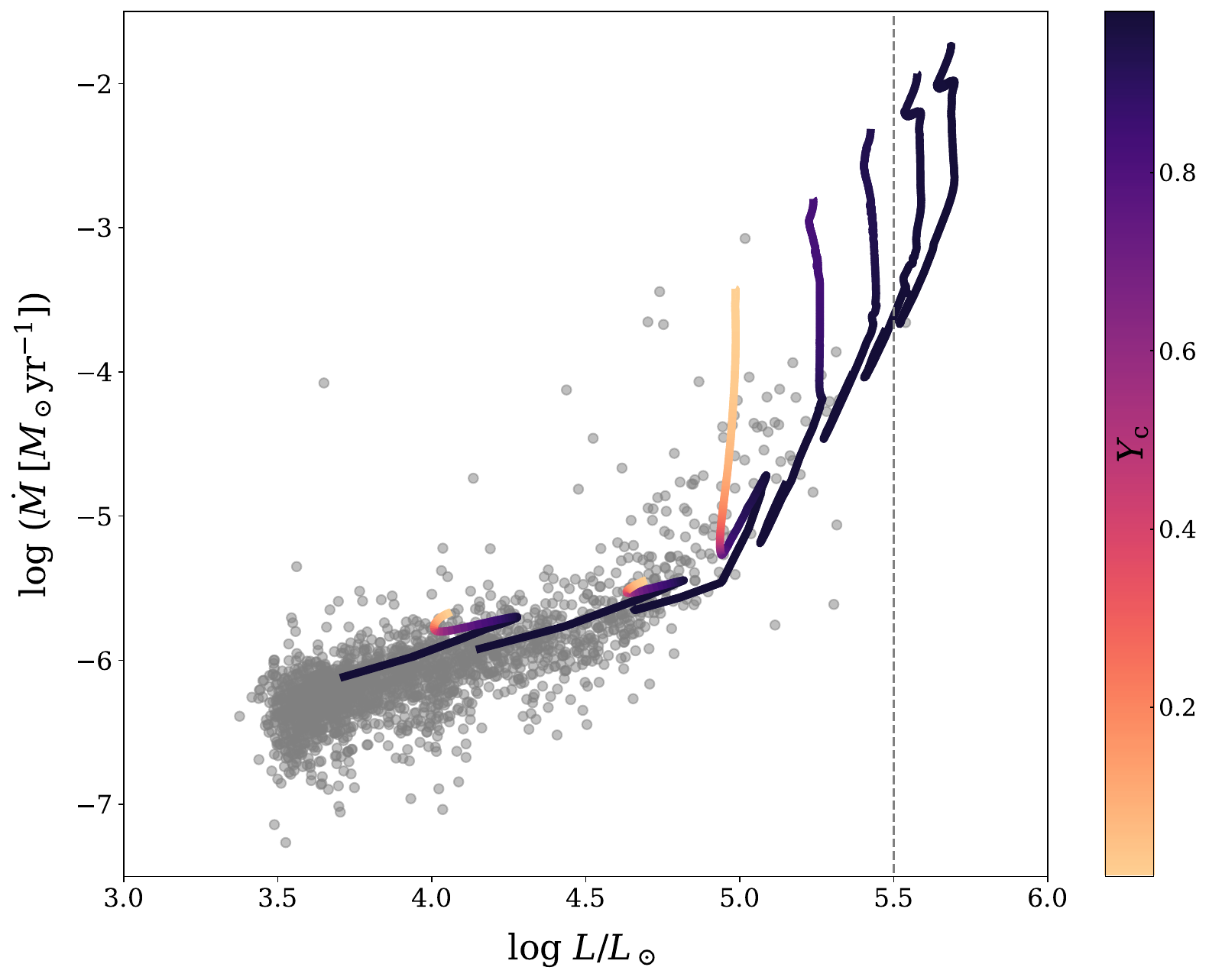}
   \caption{Mass-loss implementation into MESA models. The background stars are mass-loss rates from Yang et al. (2023). Over-plotted are the mass-loss rates versus luminosity relations from MESA evolution models from 10\,\Msun\ to 40\,\Msun\ in steps of 5\,\Msun. The colour coding is according to the central helium abundance ($Y_{\rm C}$). The dotted vertical line is the HD limit from Davies \& Beasor (2020).}
              \label{fig:impl}
    \end{figure}
    
For this reason, we parameterized the Yang et al. (2023) data as a function of both $L$ and $M$, as follows. For the shallow part below the kink, we apply:

\begin{equation}
    \log \dot{M} = -8 + 0.7  \log(L/\Lsun) - 0.7  \log(M_{\rm cur}/\Msun)
,\end{equation}
from Yang et al. (2023) data below the kink.
While for the steeper part above the kink, it is:

\begin{equation}
    \log \dot{M} = -24 + 4.77 \log(L/L_{\odot}) -3.99 \log(M_{\rm cur}/M_{\odot}),
\end{equation}
which is connected to the lower slope of the Yang et al. data, but where the coefficients are from Vink et al. (2011). We note that the 4.77 $\log L$ coefficient for the high $L/M$ case is similar to both that discussed in Beasor er al. (2020)\footnote{The independent analysis by Decin et al. (2023) has led to a corrected recipe in Beasor et al. 2023.} and the upper part of the Yang et al. relation (as shown in Fig.\,1). The $\log M$ coefficient is less well established, as stellar masses ($M_{\rm cur}$) cannot be obtained directly from RSG observations. 
However, as a larger potential well would make it harder to overcome gravity, a current mass dependence in the denominator should be considered to be physically appropriate.
We note that Beasor et al. (2020) employed initial masses characteristic for the cluster they reside in and found a linear dependence of just -0.23. However, this relies on just one data point at high mass and three at lower masses. A logarithmic dependence would yield a steeper dependence in line with Vink et al. (2011). 

We go on to implement the above mass-loss kink in MESA stellar evolution models, with
inputs similar to those in Sabhahit et al. (2022).  
One the largest uncertainties in the structure of massive stars involve mixing properties. Instead of performing a full parameter study, we only entertained a proof of concept. If our RSG mass loss concept is able to remove the hydrogen envelope, we wished to make our task most challenging by keeping mixing parameters to an absolute minimum.
We thus employed a low step-overshooting value of $\alpha_{\rm ov} = 0.1$. If we had applied a higher step overshooting value of $\alpha_{\rm ov} = 0.5$, we would have possibly removed the H envelope for lower mass-loss rates (such as those provided by Beasor et al. 2020). In other words, the key is the steep $\Gamma$-dependence (steep $L$ and inverse steep $M$) provided by both the Beasor et al. and Yang et al data.

\section{MESA stellar evolution input}

The stellar evolution models are calculated using the 1D stellar evolution code MESA (version r12115) (Paxton et al. 2011, 2013). The relevant mixing parameters are as follows: mixing length parameter, $\alpha_{\rm MLT} = 1.5$, a semi-convective diffusive efficiency of $\alpha_{\rm sc} = 1$, and a step overshooting prescription above the H core with an efficiency of $\alpha_{\rm ov} = 0.1$. The models are non-rotating to avoid rotational mixing and we also did not employe MLT$++$. The initial chemistry is $Z = 0.02, Y = 0.28, X=0.7$
The initial masses are 10, 15, 20, 25, 30, 35, and 40\,\Msun.

The mass-loss recipe during the RSG phase (with $T_{\rm eff} <$  4000 K) is given by Eqs. 1 and 2 above; for all other phases, it is taken from 
Sabhahit et al. (2022), which is a implementation of optically thin winds for O-type stars (Vink et al. 2001) and enhanced winds for high $\Gamma$ from Vink et al. (2011). In this paper, for the hydrogen-burning main sequence (MS), this implies that the canonical rates of Vink et al. (2001) are used prior to the RSG phase, but afterwards, the models generally have mass-loss rates that remain enhanced.

\section{Results}

Figure 2 has the same background stars from Yang et al. (2023) but it shows an overplotting of the mass-loss rates versus luminosity relations from MESA evolution models from 10\,\Msun\ to 40\,\Msun\ in steps of 5\,\Msun. The color coding is done according to the central helium abundance. If the color is dark, this indicates little time spent during the phase. For example, the 25\,\Msun\ model spend less than 20\% in the helium burning RSG regime before it quickly returns to the blue.

The main reason why the mass-loss runaway is so strong in the regime of RSGs is that luminosity and mass become decoupled (e.g., Farrell et al. 2020). While the initial ZAMS mass largely sets the core mass and the entry luminosity during the RSG phase, the strong mass loss leads to an ever decreasing current-day stellar mass. This is unique to supergiants and would not occur during the H-burning MS.
 
Figure 3 showcases a traditional HR diagram, with dots every 0.1 decrease in central Y ($Y_{\rm C}$) during core He burning. The models above 25\,\Msun\ have been stopped before they evolve towards the hotter Wolf-Rayet (WR) stage.

  \begin{figure}
   \centering
   \includegraphics[width=0.45\textwidth]{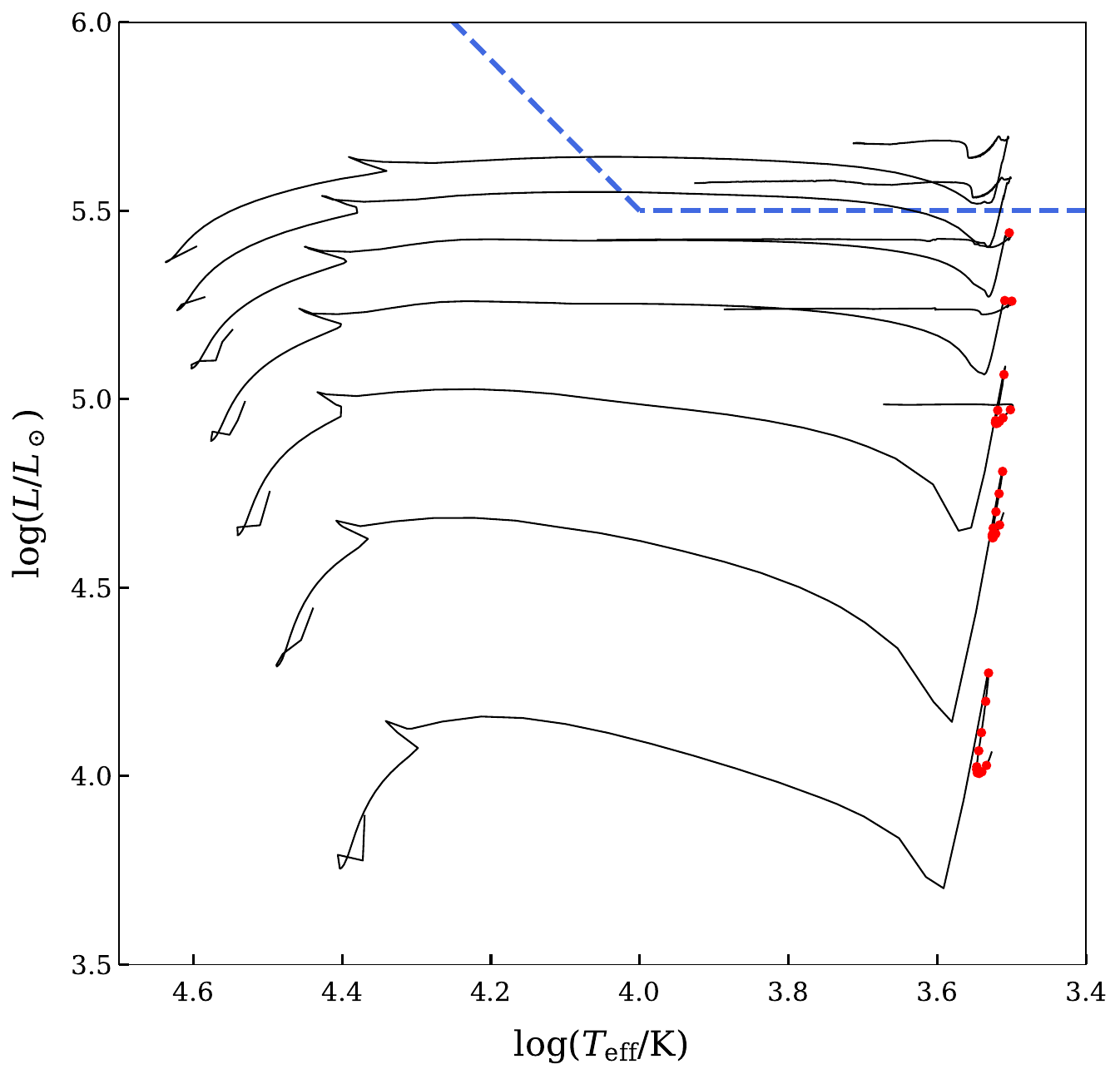}
   \caption{HR diagram with dots every 0.1 decrease in central Y during core He burning. The models above 25\,\Msun\ have been stopped before they evolve towards the WR stages. The Humphreys-Davidson limit (from Davies \& Beasor 2020) is indicated by the blue dashed line.}
              \label{fig:HRD}
    \end{figure}

Figure 3 shows that when $M$ and $L$ increase, the stars spend successively less time during the core He burning lifetime. By $M_{\rm zams} \simeq 30\,\Msun$, they naturally predict the HD limit at $\log L/\Lsun \simeq 5.5$, which has been found empirically (Davies et al. 2018; Macdonald et al. 2022). 

\section{Discussion}

We have implemented a $\Gamma$-dependent mass-loss rate for RSG stars, based on an observational kink feature in the data from Yang et al. (2023) and on a physical model that leads to a mass-loss runaway. We note that this runaway only occurs due to the $\Gamma$-dependence that involves a inverse proportionality to the current stellar mass. 

We have also shown that this implementation naturally reproduces not only the sheer existence of an HD limit, but it does so even at the correct $\log L$. Although this could be considered  a coincidence as absolute mass-loss rates are yet to be universally agreed upon. Nevertheless, it should be emphasised that we have not tweaked the mass-loss rates, but simply followed a mass-loss procedure that naturally provides the currently deemed correct upper-mass limit for RSGs.

Since many (of order 50\%) massive stars are part of close binary systems (Sana et al. 2013; Kobulnicky 2014), binary interactions could of course also  (partly) be responsible for the removal of the H-envelope in RSGs (e.g., Podsiadlowski et al. 1992; Laplace et al. 2021), but this may not result in a sharp feature in the HR diagram (Vink 2022). Instead, the tipping point is most likely related to the Eddington limit against radiation pressure (Humphreys \& Davidson 1979). 

This is not the first time that RSG mass-loss prescriptions have been analyzed. While traditional models may have employed a Reimers (1975) law, more recent modeling has generally assumed that of de Jager et al. (1988). A compilation by Mauron \& Josselin (2011) showed that the absolute rates of de Jager et al. (1988) are not out of the ordinary. However, while this oft-used prescription does include a luminosity dependence, it does not include a stellar mass dependence. 

The issue with many previous generations of single-star models was that RSG mass-loss rates have often been artificially enhanced (e.g., Meynet et al. 2015 increased mass-loss rates by factors of 10 and 25) and while these efforts were very insightful, they were also (quite correctly) disputed (e.g., Beasor et al. 2020). 

The advance of our physically motivated (with the empirical kink feature also grounded empirically) mass-loss prescription is that it naturally reproduces the upper limit for RSGs, without there being any (possibly artificial) need for
additional energy transport in super-adiabatic layers, such as via MESA's MLT$++$ parametrisation (e.g., Sabhahit et al. 2021).

Important future observations could entail similarly large RSG samples for other galaxies, such as the LMC (Groenewegen \& Sloan 2018), the Milky Way, as well as M31 and M33 (Wang et al. 2021). It would also be useful to see how dust mass loss determinations compare to gas methods (Decin 2021). 
Regarding the possible progeny of RSGs, additional future studies using flash spectroscopy of SNe would be useful to further test our proposed $\Gamma$-dependent runaway mass-loss mechanism (Gr\"afener \& Vink 2016).

Finally, as the SMC data of Yang et al. (2023) already provide such high mass-loss rates at low $Z$ (about 20\% $\Zsun$), this may suggest the RSG mass-loss rates are strongly $\Gamma$-dependent and only weakly dependent on $Z$ (at least in comparison to ZAMS OB stars). 

Future observations and modelling would need to confirm if this would result in a roughly Z-independent HD limit, with enormous implications for stripped-star SNe, black hole formation, and gravitational wave events.

\begin{acknowledgements}
We thank the anonymous referee for raising points that have helped clarify our paper. Part of this work was supported by STFC grant number ST/V000233/1. We warmly thank the MESA developers for making their stellar evolution code publicly available.

\end{acknowledgements}

\end{document}